\documentclass{article}

\usepackage{arxiv}

\usepackage[utf8]{inputenc} 
\usepackage[T1]{fontenc}    
\usepackage{hyperref}       
\usepackage{url}            
\usepackage{booktabs}       
\usepackage{amsfonts}       
\usepackage{nicefrac}       
\usepackage{microtype}      
\usepackage{lipsum}
\usepackage{graphicx}
\usepackage{amsmath}

\title{Temporal characterization of broadband femtosecond laser pulses by third-harmonic dispersion scan with ptychographic retrieval}

\author{
  Tiago Gomes$^{1,+}$\\
  IFIMUP and Departamento de Física e Astronomia\\
  Faculdade de Ciências da Universidade do Porto\\
  Rua do Campo Alegre s/n, 4169-007 Porto, Portugal \\
  \texttt{tsgomes@fc.up.pt} \\
   \And
  Miguel Canhota$^{+}$ \\
  IFIMUP and Departamento de Física e Astronomia\\
  Faculdade de Ciências da Universidade do Porto\\
  Rua do Campo Alegre s/n, 4169-007 Porto, Portugal \\
  \texttt{mcanhota@fc.up.pt} \\
   \AND
   Helder Crespo \thanks{ Present Address: Blackett Laboratory, Imperial College, London SW7 2AZ, UK}\\
  IFIMUP and Departamento de Física e Astronomia\\
  Faculdade de Ciências da Universidade do Porto\\
  Rua do Campo Alegre s/n, 4169-007 Porto, Portugal \\
   \texttt{hcrespo@fc.up.pt} 
}

\begin{document}
\maketitle

$^{1}$ Corresponding author: tsgomes@fc.up.pt

$^{+}$ These authors contributed equally to this work.

\begin{abstract}
We present a new variant of dispersion scan (d-scan) based on surface third-harmonic generation (STHG) and a ptychographic algorithm taylored for full retrieval (amplitude and phase) of broadband laser pulses. We demonstrate the technique by temporally measuring and compressing few-cycle pulses with $7 \,\text{fs}$ and $2.5 \,\text{nJ}$ from a Ti:Sapphire oscillator, using a sapphire window as the nonlinear medium. The results are in very good agreement with standard second-harmonic d-scan measurements based on a nonlinear crystal. The intrinsically broadband and phase-matching-independent nature of STHG make this technique very suitable for the characterization of ultrashort laser pulses over a broad wavelength range extending into the mid-infrared.
\end{abstract}


Broadband few-cycle laser pulses have numerous applications in ultrafast science, from the probing of ultrafast molecular dynamics \cite{Probing_molecular} to nonlinear bioimaging \cite{Maibohm:19} and controlling electron emission in nanotips with attosecond resolution \cite{hommelhoff}.
However, a complete study of ultrafast dynamics requires full knowledge of the complete electric field of the ultrashort pulses used to excite and probe these dynamics. Therefore, several techniques have been developed over the years to temporally characterize ultrashort pulses, such as FROG \cite{FROG}, SPIDER \cite{SPIDER} and MIIPS \cite{MIIPS}. These techniques, although adequate, usually require complex setups involving interferometers or pulse shapers. 

A more recent technique is dispersion scan (d-scan) \cite{dscan1,dscan2}, which enables the full temporal characterization of ultrashort pulses using a simple and fully inline setup. In this method, a nonlinear signal is measured for different amounts of dispersion applied to a light pulse, creating a 2D d-scan trace which, through optimization algorithms \cite{dscan2,GP_Dscan,DE_Dscan,COPRA,Kleinert:19}, allows reconstructing the spectral amplitude and/or phase and obtaining the exact temporal profile of the ultrashort pulse. The fact that d-scan is a single beam technique that does not require crossing of different beams within a nonlinear medium results in ease of alignment and avoidance of beam-smearing artifacts that would otherwise distort the measurement of few-cycle pulses.

The most common nonlinear signal used for d-scan is second-harmonic generation (SHG), usually produced in a nonlinear crystal such as BBO. However, using SHG may be a problem when dealing with octave-spanning pulses, due to the spectral overlap between the fundamental and the SHG beams, although certain implementations of d-scan can actually take advantage of this overlap to measure and stabilize the carrier-envelope phase (CEP) \cite{Octave_spanning} or to access the absolute CEP of the pulses in situ \cite{CEP_dscan}. Third-harmonic generation (THG) d-scan has been reported \cite{silvaMidIRUltrabroadbandThird2013a,THG_absorbing,this_dscan}, namely using graphene as a nonlinear medium \cite{silvaMidIRUltrabroadbandThird2013a} to take advantage of its extraordinarily high third-order nonlinear optical susceptibility. However, laser-induced damage to graphene samples \cite{Damage_Graphene} could limit the possibility of using graphene as a universal material to characterize ultrashort pulses. As an alternative route, STHG can overcome this problem \cite{THG_interfaces,THG_FHG_interfaces}. Although having a relatively low efficiency - definitively lower than that of graphene - it occurs at all interfaces and can be used to fully characterize the temporal profile of ultrashort pulses with arbitrary bandwidth and center frequency, limited only by the detection process. For example, a standard spectrometer with a silicon-based detector should enable d-scan measurements of pulses with center frequencies up to $3\,\mu \text{m}$. STHG-based FROG has been demonstrated with $\approx 100 \,\text{fs}$ pulses \cite{STHG_FROG} and more recently with sub-10-fs pulses in the visible \cite{STHG_FROG1} and near-infrared \cite{STHG_FROG2} regions. However, to the best of our knowledge, no STHG d-scan measurement has been reported so far.

In this Letter we have successfully applied STHG to determine the temporal profile of broadband $7\,\text{fs}$ laser pulses from a Ti:Sapphire oscillator by means of a d-scan measurement. The obtained results are in very good agreement with standard SHG d-scan using a BBO crystal, showing that STHG d-scan is a powerful and universal tool to characterize ultrashort pulses.

\begin{figure}[htbp]
\centering
\fbox{\includegraphics[width=0.7\linewidth]{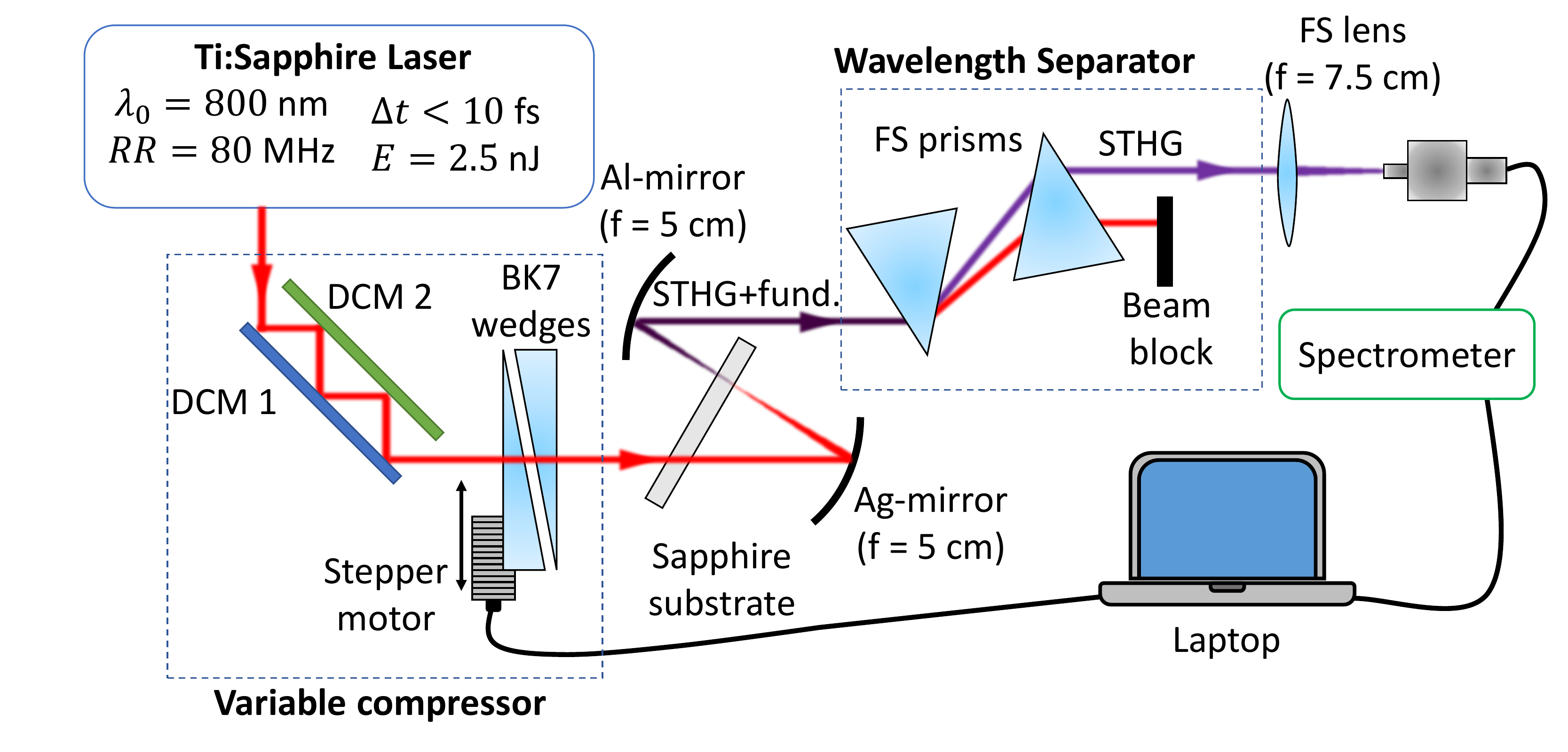}}
\caption{STHG d-scan experimental setup. DCM1,2: double- chirped mirrors (see text for more details).}
\label{fig:THG_d_setup}
\end{figure}

The experimental setup for STHG d-scan is given in Fig.\,\ref{fig:THG_d_setup}. In our experiments, we used an ultra-broadband Ti:Sapphire laser oscillator (Femtolasers Rainbow CEP) delivering pulses with a central wavelength of $800\,\text{nm}$, sub-10-fs duration, $2.5\,\text{nJ}$ of energy and a repetition rate of $80\,\text{MHz}$. Like in most d-scan implementations, the setup comprises three sections: a variable compressor (usually already in place and responsible for adjusting the total dispersion applied to the laser pulse) a nonlinear signal generator (where the STHG is performed) and a spectral measurement. In the variable compressor, the beam undergoes 8 bounces off ultra-broadband double-chirped mirrors (Laser Quantum Ltd) that introduce negative dispersion before crossing a pair of BK7-glass wedges for variable dispersion compensation. One of the wedges is connected to a motorized stage for scanning and fine-tuning the applied dispersion. The pulse energy is adjusted with a variable neutral-density filter prior to the nonlinear section, where the beam crosses a $0.5\,\text{mm}$ thick sapphire substrate and is focused back on the same substrate with a concave silver-coated spherical mirror ($\text{f}=5\,\text{cm}$) at a incidence angle of $19^{\circ}$ (a smaller angle would have been preferable to minimize astigmatism, but this was the best compromise in our setup given our beam size and the useful aperture of the components). The resulting STHG signal is then collimated with an aluminum-coated concave spherical mirror with the same focal length. Spherical mirrors were used instead of lenses to reduce chromatic aberration. Note that high-quality, low scatter off-axis parabolic mirrors could have been used instead. We opted for spherical mirrors at a relatively low incidence angle due to their better quality focused spot and higher focused intensity compared to standard off-axis parabolic mirrors. 

\begin{figure}[htbp]
\centering
    \fbox{\includegraphics[width=0.7\linewidth]{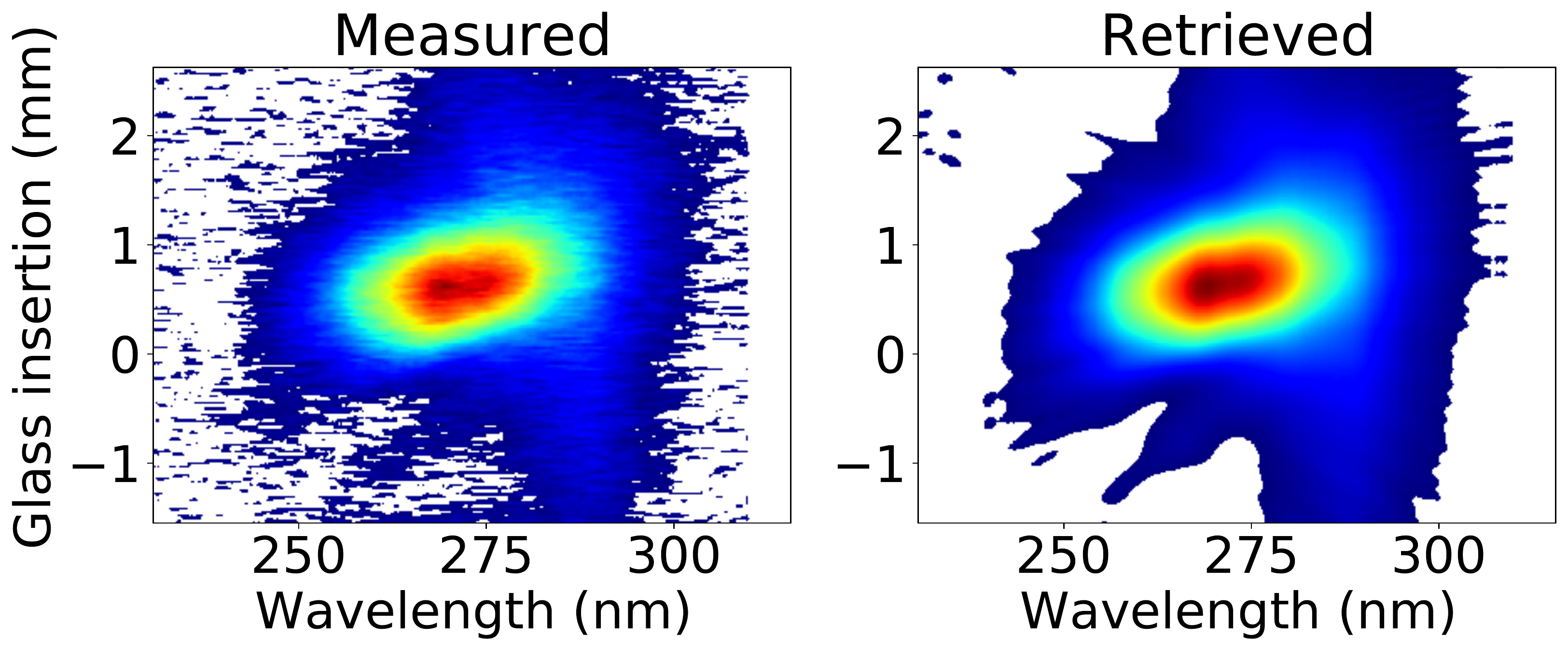}}
\caption{Measured (left) and retrieved (right) STHG d-scan using a $0.5 \,\text{mm}$-thick sapphire substrate.}
\label{fig:THG_d_scan}
\end{figure}

The collinear fundamental and STHG beams are sent through a wavelength separator composed of a set of two prisms, to spatially separate the two beams prior to the spectral measurement \cite{silvaMidIRUltrabroadbandThird2013a}. A double-pass scheme was used to remove spatial chirp (for clarity, only one pass is shown in Fig.\,\ref{fig:THG_d_setup}). The spectrum of the STHG signal was then recorded as a function of dispersion with a fiber-coupled spectrometer (Ocean Optics HR4000) to obtain the measured STHG d-scan trace (Fig.\,\ref{fig:THG_d_scan}).

\begin{figure}[htbp]
\centering
\fbox{\includegraphics[width=0.7\linewidth]{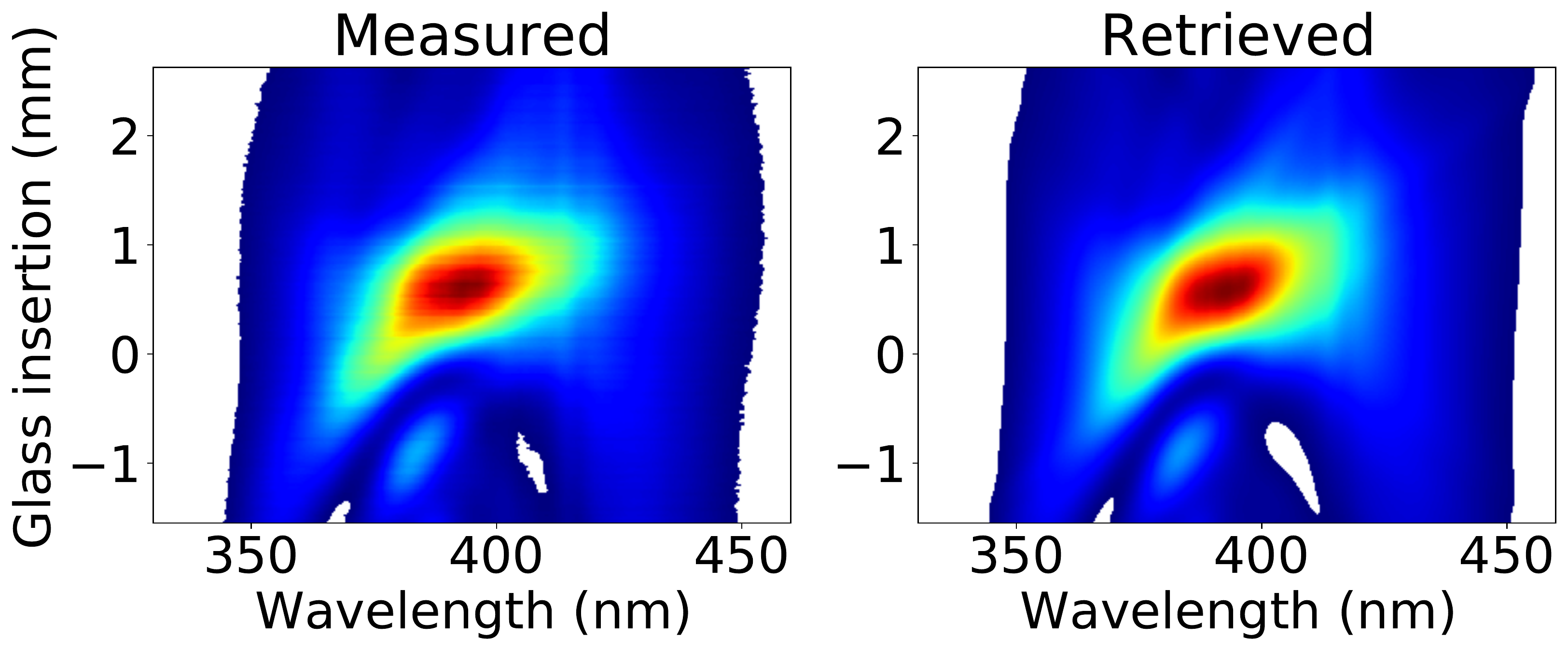}}
\caption{Measured (left) and retrieved (right) SHG d-scan using a $20\,\mu  \text{m}$-thick BBO crystal.}
\label{fig:SHG_d_scan}
\end{figure}

The same setup was then adapted to perform a standard SHG d-scan measurement of the pulses (Fig.\,\ref{fig:SHG_d_scan}), where we placed an off-axis parabolic mirror after the first pass in the sapphire substrate to focus the beam onto a $20\,\mu \text{m}$-thick BBO crystal thus generating the second-harmonic. A blue filter was used to eliminate the fundamental beam, and the SHG beam was then focused onto the same fiber-coupled spectrometer. We scanned the dispersion in 150 steps over a 4-mm wedge insertion range for both techniques, using an integration time of 1\,s for the STHG d-scan and 10\,ms for the SHG d-scan.

The retrieval of the electric field of the pulses follows the recently developed ptychographic approach to pulse retrieval from 2D-traces \cite{Sidorenko:16, Wilhelm:20}, which shows promise for retrieving complex pulses from incomplete traces and/or traces with low signal-to-noise ratios \cite{Sidorenko:16} and which we have adapted in order to extend its applicability to the important case of broadband few-cycle pulses. Unlike most pulse retrieval algorithms, ptychographic retrieval does not rely on the minimization of a merit function (e.g. the d-scan error $G$ \cite{dscan1}); instead, it iteratively approximates the field by combining the field from a previous iteration with the measured and the calculated d-scan traces via an adequate transfer function, as detailed below.

The ptychographic d-scan algorithm \cite{Wilhelm:20} starts by applying the variable phase of the compressor (e.g., from a wedge pair), $\phi_{k}(\omega)$, to a first guess of the field of the pulse, $E_{k}(\omega)$, obtaining

\begin{equation}
    \centering
    X_{k}(\omega)=E_{k}(\omega)e^{i\phi_{k}(\omega)},
\end{equation}
where $k$ denotes a given position of the variable compressor (e.g., a given wedge insertion). For the case of the second-harmonic, the nonlinear effect is modeled as the square of the signal in the time domain, i.e.,
\begin{equation}
    \centering
    \psi_{k}(t)=\left(\mathcal{F}^{-1}\left\{ X_{k}(\omega)\right\} \right)^{2},
\end{equation}

while for the case of the surface third-harmonic, the nonlinear effect is modeled by

\begin{equation}
    \centering
    \psi_{k}(t)=\left(\mathcal{F}^{-1}\left\{ X_{k}(\omega)\right\} \right)^{3}.
\end{equation}

The corresponding complex spectral amplitude is given by

\begin{equation}
    \centering
    \Psi_{k}(\omega)=\mathcal{F}\left\{ \psi_{k}(t)\right\}
\end{equation}
and the simulated d-scan trace, $T_k^{\text{sim}}(\omega)$, can be modeled as

\begin{equation}
    \centering
    T_{k}^{\text{sim}}(\omega)=|\Psi_k(\omega)|^2.
\end{equation}

Following the ptychographic approach, a modulus constraint is applied,

\begin{equation}
    \centering
    \Psi'(\omega)=\sqrt{T_{k}(\omega)}\frac{\Psi_{k}(\omega)}{\left|\Psi_{k}(\omega)\right|},
\end{equation}

where $T_{k}(\omega)$ is the normalized measured d-scan trace. A new signal, $\psi'_{k}(t)$,
is then determined by

\begin{equation}
    \centering
    \psi_{k}'(t)=\mathcal{F}^{-1}\left\{ \Psi'(\omega)\right\}. 
\end{equation}

In order to obtain the new (iterated) field, a transfer function is computed, which for the second-harmonic reads

\begin{equation}
    \centering
    X'_{k}(t)=\frac{1}{2}\left[2X_{k}(t)+\frac{X_{k}^{*}(t)}{\left|X_{k}(t)\right|_{max}^{2}}\left(\psi_{k}'(t)-\psi_{k}(t)\right)\right].
\end{equation}

For the STHG process, the above expression is substituted by

\begin{equation}
    \centering
    X'_{k}(t)=\frac{1}{3}\left[3X_{k}(t)+\frac{\left(X_{k}^{*}(t)\right)^{2}}{\left|X_{k}(t)\right|_{max}^{4}}\left(\psi_{k}'(t)-\psi_{k}(t)\right)\right].
\end{equation}
These transfer functions are inspired by Newton's method for computing the square and cubic roots, respectively.

Going back to the frequency domain, $X'_{k}(\omega) = \mathcal{F}\left(X'_{k}(t)\right)$, the phase of the compressor is removed,

\begin{equation}
    \centering
    E_{k}'(\omega)=X'_{k}(\omega)e^{-i\phi_{k}(\omega)}.
\end{equation}

Finally, the field is calculated as the arithmetic average of the previous fields over all insertions, i.e.,

\begin{equation}
    \centering
    E(\omega)=\frac{\sum_{k=1}^{N}E_{k}'(\omega)}{N}.
\end{equation}

\begin{figure}[htbp]
\centering
\fbox{\includegraphics[width=0.7\linewidth]{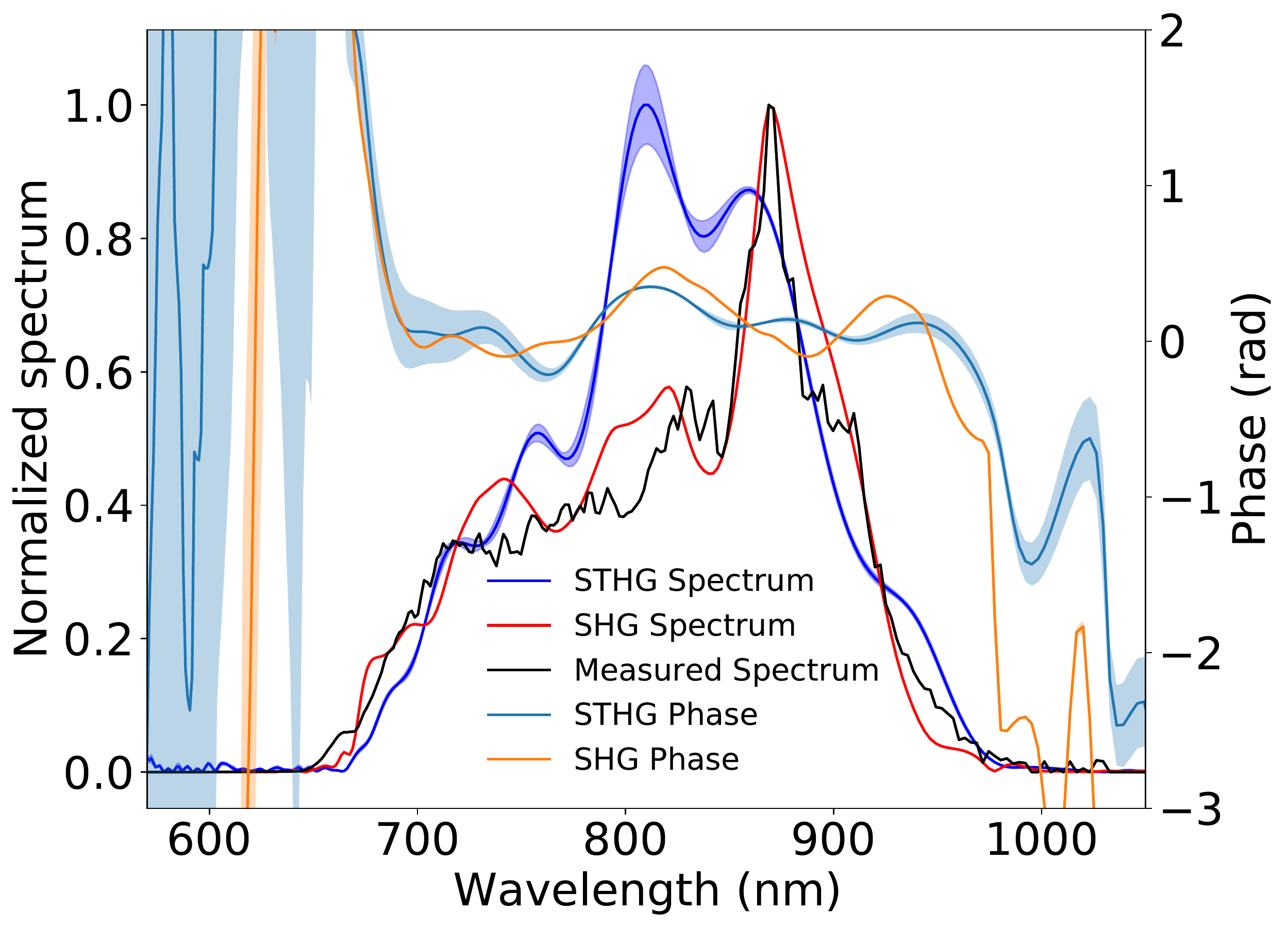}}
\caption{Mean retrieved spectral phase and amplitude for SHG d-scan (orange and red, respectively) and STHG d-scan (light blue and dark blue, respectively), along with the calculated standard deviation (shaded area). The black curve is an independent measurement of the spectral power of the laser.}
\label{fig:Phase_comp}
\end{figure}

This algorithm successfully retrieved both the spectrum and the phase of relatively long (40\,fs) pulses derived from a grating compressor \cite{Wilhelm:20}.
The good agreement between the experimental and retrieved spectrum reported in the initial work is not observed in our present case of shorter, few-cycle pulses.
We therefore introduced a new extension to the algorithm in order to improve the accuracy of the retrieved spectrum, to be called the spectrum and phase approach (SPA).
We observed numerically that the first run of the algorithm (first stage) yields a retrieved spectrum that can be significantly different from the directly measured spectrum, which can be attributed to the fact that the generation of the nonlinear effect is assumed to be ideal. With the retrieved trace and the acquired trace, it is possible to calculate a spectral filter (Eq.\,4 in \cite{dscan1}).
A new trace can be constructed by factoring out the filter from the measured trace. With this new modified trace the algorithm is run again (second stage), where the retrieved spectrum is now close to the measured one.
The SPA was applied to the experimental data, where the measured traces were not calibrated in intensity. Despite the lack of intensity calibration, the retrieved spectra are faithful to the measured spectrum, particularly in the case of the SHG d-scan measurement (Fig. \ref{fig:Phase_comp}). We observed that the spectral phases retrieved by both techniques (with and without SPA) are nevertheless very similar in the region defined by the laser spectrum, since phase retrieval by d-scan is not corrupted by nonlinear conversion bandwidth issues.

All traces are 150 $\times$ 384 pixels in size (insertion $\times$ frequency). One individual retrieval process comprises the two stages, including the SPA treatment, and takes around 4 seconds in the SHG variant and 5 seconds in the STHG variant, using a 2019 AMD Ryzen 7 processor. The stopping criteria for each stage is that the difference between consecutive simulated traces be less than $8\times 10^{-4}$ percent. For differences smaller than the one referred, some oscillations appeared in the spectrum, and therefore the stopping criteria was adjusted to overcome this problem.

The measured and the retrieved STHG d-scan using a sapphire plate can be seen in Fig.\,\ref{fig:THG_d_scan}. The retrieved d-scan trace shown corresponds to one out of 10 retrievals, performed for different initial guesses (Gaussian spectra with random second- and third-order spectral phases). The retrieved spectral phase and intensity used to reconstruct the pulse's temporal profile are the mean retrieved spectral phase and intensity of all retrievals. 

The mean d-scan error is $8.1\%$, which is relatively high due to the fact that the retrieval algorithm does not retrieve the noise that appears on the d-scan trace. However, we were able to reconstruct the main features of the STHG d-scan, namely the trace's small tilt, caused by residual third-order dispersion, and the small features that appear around the trace's center of mass.

The same retrieval process was applied to the measured SHG d-scan, whose measured and retrieved traces can be seen in Fig.\,\ref{fig:SHG_d_scan}.

In this case, the mean d-scan error G is equal to $5.7\,\text{\%}$, and all the features were correctly reproduced by the retrieval. Visually, both retrieved d-scan traces are very similar to the measured ones. The retrieved spectral phases and amplitudes obtained from both techniques are shown in Fig.\,\ref{fig:Phase_comp}.

We see that in the region with higher spectral power the retrieved spectral phases using both techniques are in very good agreement. Both spectral phases diverge rapidly for wavelengths lower than $\approx$\,680\,nm and are also relatively flat over the most significant spectral region ($\approx$\,680-960\,nm).

The spectral intensity retrieved using SHG d-scan is in excellent agreement with the measured one. However, the retrieved spectral intensity with the STHG d-scan has significant differences compared to the measured one, which we attribute to the presence of astigmatism induced by the silver-coated concave mirror. This effect was visible as small shifts in the nonlinear signal frequencies that depend on the angle and distance between the laser beam and the sapphire plate.

\begin{figure}[htbp]
\centering
\fbox{\includegraphics[width=0.7\linewidth]{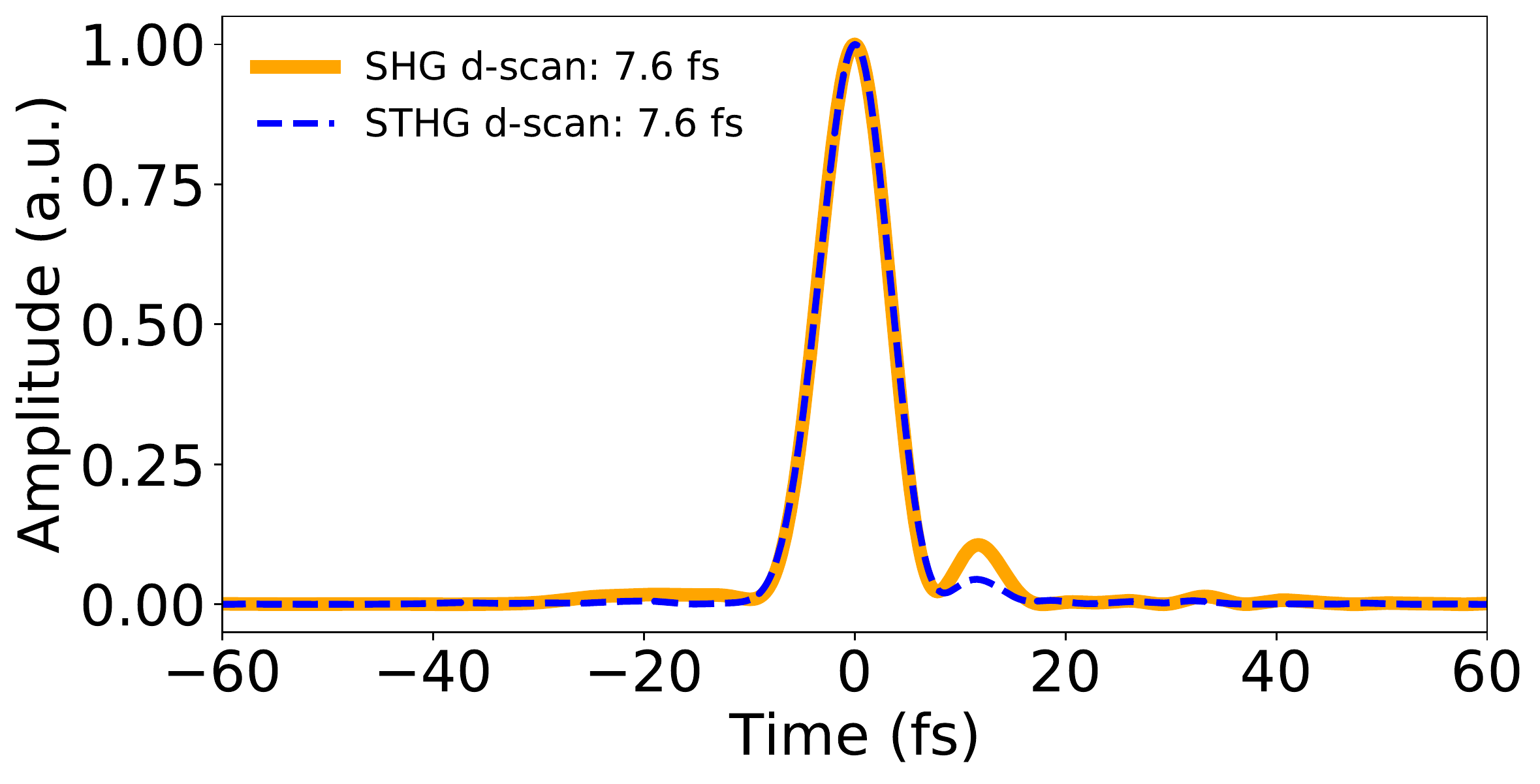}}
\caption{Pulse temporal profile retrieved by SHG d-scan (orange) and STHG d-scan (dashed blue).}
\label{fig:Temp_comp}
\end{figure}

The temporal profile of the ultrashort pulse retrieved by the two techniques is given in Fig.\,\ref{fig:Temp_comp}.

The agreement between both retrievals is very good, with a full-width-at-half maximum (FWHM) pulse duration of $7.6 \pm 0.3\,\text{fs}$ (STHG d-scan) and $7.6 \pm 0.1\,\text{fs}$ (SHG d-scan). If the pulses are plotted using the measured spectrum instead of the retrieved spectra, they are virtually indistinguishable by eye, which again confirms the robustness of the phase retrieval. In both cases, we observe a small post-pulse, synonym of residual third-order dispersion, whose position is identical in both cases. These results are also in good agreement with previous published results \cite{dscan1} using the same laser source.


In conclusion, we have demonstrated surface THG d-scan measurements of few-cycle lsaer pulses using a $0.5 \,\text{mm}$-thick sapphire substrate as the nonlinear medium and a ptychographic retrieval algorithm adapted to broadband pulses by introducing a second retrieval step to account for the non-ideal nature of the nonlinearity (SPA approach) and fully retrieve the pulses (spectrum and phase) without the need to calibrate the measured nonlinear signal in intensity. These results are in very good agreement with the established technique of SHG d-scan based on BBO crystals, showing that surface THG d-scan is a powerful technique for the simultaneous characterization and compression of ultrashort pulses. The intrinsic broadband nature of STHG should allow the full characterization of ultrashort pulses with arbitrary frequencies and bandwidth via d-scan, only limited by the detection method.
Although SPA was applied to the ptychographic algorithm, it is not exclusive to it, and could be used in other full-field d-scan retrieval algorithms.

\bibliographystyle{unsrt}  
\bibliography{references}  


\end{document}